\documentclass[aps,pre,reprint,amsmath,amssymb,floats,longbibliography]{revtex4-2}

\usepackage{graphicx}
\usepackage{bm}
\usepackage{newunicodechar}
\usepackage[utf8]{inputenc}
\usepackage[T1]{fontenc}
\newunicodechar{ȩ}{\k{e}}
\usepackage{color}

\begin{document}
\title{Eigenstate thermalization hypothesis in two-dimensional {\em XXZ} model with or
without SU(2) symmetry}

\author{Jae Dong Noh}
\affiliation{Department of Physics, University of Seoul, Seoul 02504, Korea}
\date{\today}

\begin{abstract}
    We investigate the eigenstate thermalization properties of the spin-1/2
    {\em XXZ} model in 
    two-dimensional rectangular lattices of size $L_1\times L_2$ 
    under periodic boundary conditions. Exploiting the symmetry
    property, we can perform an exact 
    diagonalization study of the energy eigenvalues up to
    system size $4\times 7$ and of the energy eigenstates up to $4\times 6$.
    Numerical analysis of the Hamiltonian eigenvalue spectrum and 
    matrix elements of an observable in the Hamiltonian eigenstate basis 
    supports that the
    two-dimensional {\em XXZ} model follows the eigenstate thermalization
    hypothesis. When the spin interaction is isotropic 
    the {\em XXZ} model Hamiltonian conserves the total spin and has 
    SU(2) symmetry. We show that the eigenstate thermalization
    hypothesis is still valid within each subspace where the total spin
    is a good quantum number.
\end{abstract}
\maketitle

\section{Introduction}\label{sec:1}
The eigenstate thermalization hypothesis~(ETH) explains 
the mechanism for thermalization of isolated quantum
systems~\cite{Deutsch.1991,Srednicki.1994}. The ETH guarantees that a
quantum mechanical expectation value of a local observable relaxes to the
equilibrium ensemble averaged value and fluctuations in the steady state
satisfy the fluctuation dissipation theorem~(see Ref.~\cite{D'Alessio.2016}
and references therein).

Numerous studies have been performed to test validity of the ETH 
since the early work of Ref.~\cite{Rigol.2008}.
The spin-1/2 {\em XXZ}
model~\cite{Rigol.2009,Steinigeweg.2013,Kim.2014,Alba.2015,Jansen.2019,Noh.2019,Brenes.2020,Noh.2020,Noh.2021,Schonle.2021}
and the quantum Ising spin
model~\cite{Kim.2014,Fratus.2015,Mondaini.2016,Mondaini.2017,Dymarsky.2018} are the
paradigmatic model systems for ETH study. 
The {\em XXZ} model is useful since it describes a hardcore boson
system which is relevant to experimental ultracold atom
systems~\cite{Trotzky.2012,Kaufman.2016,Orioli.2018,Tang.2018,Langen.2015}.
Moreover, integrability in these models can be tuned easily in
one-dimensional lattices.  A thermal/nonthermal behavior and  
a crossover between them have been studied comprehensively using the model
systems~\cite{Rabson.2004,Rigol.2009,Santos.2010h1,Cassidy.2011,Steinigeweg.2013,Modak.2014,Kim.2014,Vidmar.2016,Brenes.2020,Brenes.2020vi,Noh.2021ykg}.

The ETH has been examined mostly in one-dimensional spin systems, and
there are only a few works for two-dimensional 
systems~\cite{Rigol.2008,Mondaini.2016,Mondaini.2017,Fratus.2015,Lan.2017}. 
In this work, we study the eigenstate thermalization property of the
spin-1/2 {\em XXZ} model in two-dimensional rectangular lattices. In comparison
with the Ising spin systems~\cite{Fratus.2015, Mondaini.2016, Mondaini.2017}, 
the {\em XXZ} model is characterized by the conservation of 
the magnetization in the $z$ direction. Furthermore,
it possesses the SU(2) symmetry when the spin interaction is 
isotropic~\cite{Halpern.2020}. 
The SU(2) symmetry conserves the magnetization in all directions, but the
total spin operators in different directions do not commute with each other. 
Such a non-Abelian symmetry has a nontrivial effect on many-body
localization~\cite{Potter.2016,Protopopov.2017}, quantum
thermalization~\cite{Halpern.2016,Halpern.2020,Kranzl.2022}, 
and entanglement entropy~\cite{Majidy.2023}.

This paper is organized as follows. In Sec.~\ref{sec:2}, we introduce 
the {\em XXZ} Hamiltonian with nearest and next nearest neighbor interactions in 
two-dimensional rectangular lattices. The symmetry property of the
Hamiltonian is summarized.
In Secs.~\ref{sec:3} and \ref{sec:su2}, we present results of a numerical 
exact diagonalization study. First, we will show in Sec.~\ref{sec:3}
that the ETH is valid in the {\em XXZ} model without SU(2) symmetry. 
In Sec.~\ref{sec:su2}, we proceed to show that the
SU(2) symmetric {\em XXZ} model also satisfies the ETH in each SU(2) subsector.
Our work extends the validity of the ETH to the two-dimensional {\em XXZ}
model. 

\section{Two-dimensional {\em XXZ} model}\label{sec:2}
\begin{figure}
    \includegraphics*[width=\columnwidth]{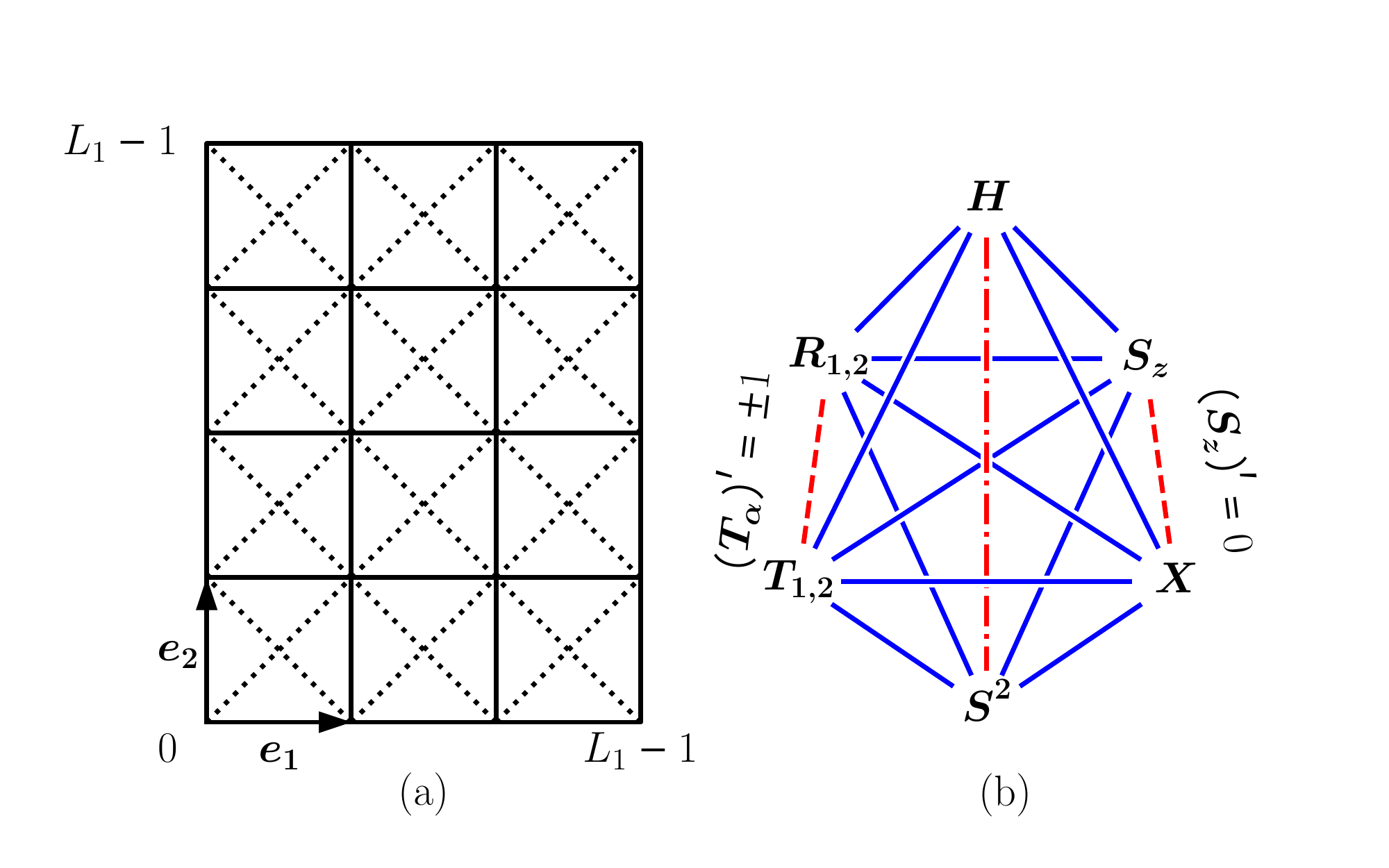}
    \caption{(a) Rectangular lattice of size $L_1\times L_2$ 
        under periodic
        boundary conditions in the horizontal~($\bm{e}_1$) and
    vertical~($\bm{e}_2$) directions. (b) Commutation relations among the 
    {\em XXZ} Hamiltonian and symmetry operators.
    Mutually commuting operators are connected with a solid line. 
    A dashed line connect operators which are commuting only within the
    subspace with specific quantum numbers of the symmetry operator.
    The Hamiltonian and $\bm{S}^2$, connected by a dashed-dotted line,
    commutes only when $\Delta = 1$.
}
\label{fig1}
\end{figure}

We consider the spin-1/2 {\em XXZ} model on a two-dimensional rectangular
lattice. The Pauli spin $\bm{\sigma}_{\bm r} = (\sigma^x_{\bm r},
\sigma^y_{\bm r}, \sigma^z_{\bm r})$ resides on a lattice 
site ${\bm r}$ and the Hamiltonian is given by
\begin{equation}
    H = \lambda \sum_{\langle \bm{r}, \bm{r'}\rangle} 
        h(\bm{\sigma}_{\bm r},\bm{\sigma_{\bm r'}}) + 
        (1-\lambda) \sum_{ [\bm{r}, \bm{r'} ]} 
    h(\bm{\sigma}_{\bm r},\bm{\sigma_{\bm r'}}) ,
    \label{hamiltonian}
\end{equation}
where $\langle \bm{r}, \bm{r'} \rangle$ and $[\bm{r}, \bm{r'}]$ 
denote the pair of nearest neighbor~(nn) sites, connected by solid lines in
Fig.~\ref{fig1}(a), and of next nearest neighbor~(nnn) sites, connected by dotted
lines in Fig.~\ref{fig1}(a), respectively, and
$h(\bm{\sigma},\bm{\sigma_{r'}})$ denotes the {\em XXZ} coupling given by
\begin{equation}
    h(\bm{\sigma_r}, \bm{\sigma_{r'}}) = -\frac{J}{2} \left(
    \sigma^x_{\bm r} \sigma^x_{\bm r'}  +
    \sigma^y_{\bm r} \sigma^y_{\bm r'}  +
\Delta \sigma^z_{\bm r} \sigma^z_{\bm r'} \right) .
    \label{h_coupling}
\end{equation}
The model is defined by three parameters $J$, $\Delta$, and $\lambda$:
$\lambda$ controls the relative strength of the nn and nnn couplings, 
$\Delta$ is an anisotropy parameter, and $J$ sets the overall
energy scale which will be kept to be 1. We assume periodic boundary
conditions, 
$\bm\sigma_{  {\bm r}+ L_1 {\bm e}_1 } = \bm\sigma_{ {\bm r}+ L_2{\bm
e_2}} = \bm\sigma_{ \bm r}$ where $\bm{e}_1$ and $\bm{e}_2$ are the unit vectors
in the horizontal and vertical directions, respectively~[see
Fig.~\ref{fig1}(a)]. The {\em XXZ} coupling with $J=1$ can be rewritten as
\begin{equation}
    h(\bm{\sigma_r}, \bm{\sigma_{r'}}) = - \left(
        \sigma^+_{\bm r} \sigma^-_{\bm r'} + \sigma^-_{\bm r} \sigma^+_{\bm
    r'} + \frac{\Delta}{2} \sigma^z_{\bm r} \sigma^z_{\bm r'}\right) 
    \label{h+-}
\end{equation}
with the raising and lowering operators $\sigma^{\pm} \equiv (\sigma^x \pm
i\sigma_y)/2$. Throughout the paper, we will set $\hbar=1$.
The total number of sites will be denoted by $N=L_1 L_2$. In this work, we
only consider the lattices with even $N$.

The {\em XXZ} Hamiltonian commutes with several symmetry operators. First, 
the Hamiltonian commutes with the magnetization operator in the $z$ 
direction 
\begin{equation}
    S_z = \frac{1}{2}\sum_{\bm r}\sigma^z_{\bm r}.
    \label{Mz}
\end{equation}
The Hamiltonian also commutes with the shift operator $T_{\alpha}$
which shifts a spin state by the unit distance in the direction 
${\bm e}_\alpha$ with $\alpha=1,2$:
\begin{equation}
    T^{-1}_{\alpha} \bm{\sigma_r} T_{\alpha} = 
\bm{\sigma}_{ {\bm r}+{\bm e}_\alpha } \quad (\alpha  = {1}, {2} ).
    \label{shift}
\end{equation}
The system has the spatial inversion symmetry so that $H$ commutes with
$R_{\alpha}$ which maps a site $\bm{r}=(x,y)$ to $(-x,y)$ for ${\alpha}=1
$ or to $(x,-y)$ for ${\alpha} = 2$. Finally, the system is
invariant under the spin flip $\sigma^z \to -\sigma^z$ which is generated by
the symmetry operator $X = \prod_{ {\bm r}} \sigma^x_{\bm{r}}$.

The commutation relations are summarized by a diagram 
in Fig.~\ref{fig1}(b).~(A similar diagram for the
one-dimensional system is found in Ref.~\cite{Jung.2020}.) Note that $[X,
S_z] \neq 0$ and $[R_{\alpha}, T_{\alpha}] \neq 0$ in general. 
Thus, one cannot construct a simultaneous basis set for all the symmetry
operators. 
On the other hand, one can show that $[R_{\alpha},T_{\alpha}]|\psi\rangle=0$  
if a state $|\psi\rangle$ is an eigenstate of
$T_{\alpha}$ of eigenvalue $(T_{\alpha})'=\pm 1$. 
It implies that
the two operators commute within the subspace of 
the eigenstates of $T_{\alpha}$ with 
eigenvalues $\pm 1$, Likewise, $[X,S_z]=0$ within the subspace 
of the eigenstates of $S_z$ with eigenvalue $(S_z)'=0$.
In this work, we focus on the symmetry sector consisting of the eigenvalues of
the symmetry operators with the eigenvalues $(T_{\alpha})' = (R_{\alpha})' =
(X)'=1$ and $(S_z)'=0$, which will be referred to as the maximum symmetry
sector~(MSS).

When the spin-spin interaction is isotropic~($\Delta=1$), the Hamiltonian
is invariant under spin rotation~[SU(2) symmetry].
Consequently, each component of the total spin
$\bm{S} = \frac{1}{2} \sum_{\bm r} \bm{\sigma_r}$ 
is conserved and $\bm{S}^2 = \bm{S}\cdot\bm{S}$ becomes the symmetry
operator commuting with the Hamiltonian and 
all the other symmetry operators. The maximum
symmetry sector is then further decomposed into subsectors characterized
with the eigenvalue of $\bm{S}^2$, $(\bm{S}^2)' = s (s+1)$ with integer $s$.
The SU(2) symmetry will be investigated in detail in Sec.~\ref{sec:su2}.

We have performed the exact diagonalization study. The basis states, 
which are simultaneous eigenstates of the symmetry
operators appearing in Fig.~\ref{fig1}(b) in the MSS, can be easily 
constructed using the methods summarized in
Refs.~\cite{Baerwinkel.2000,Schnack.2008,Schnalle.2010,Heitmann.2019,Sandvik.2010,Jung.2020}.
The Hilbert space dimensionalities of the MSS are 
$D = 26$, $1392$, $15578$, and 
$183926$ when $L_1\times L_2 = 4\times 3$, $4\times5$, $4\times 6$, $4\times
7$, respectively.
When $L_1=L_2$, the system has an addition symmetry under the spatial 
rotation by a multiple of $\pi/2$. 
It will not be addressed since we only consider the lattices
with $L_1 \neq L_2$.

An energy eigenstate and a corresponding eigenvalue of $H$
in the MSS will be denoted as $|E_n\rangle$ and $E_n$, respectively, 
where the quantum
number $n=0, \cdots, D-1$ is assigned in ascending order of the energy
eigenvalue.
We will study the Hamiltonian spectrum and the matrix elements of the
observable,
\begin{equation}
    \begin{split}
        O^Z &= \frac{1}{N} \sum_{\bm r} 
    \sum_{\alpha = 1,2} \sigma_{\bm{r}}^z
\sigma_{\bm{r}+\bm{e}_\alpha}^z\\
O^J &= \frac{1}{N} \sum_{\bm r} 
        \sum_{{\alpha} = 1,2} \left( \sigma_{\bm{r}}^+
            \sigma_{\bm{r}+\bm{e}_{\alpha}}^- + 
        \sigma_{\bm{r}+\bm{e}_{\alpha}}^+ \sigma_{\bm{r}}^- \right) \\
        O^P &= \frac{1}{N} \sum_{\bm{r},\bm{r'}} 
        \sigma_{\bm{r}}^+ \sigma_{\bm{r'}}^- \\
        O^F & = \frac{1}{N} \sum_{p} \sigma_{p_1}^z \sigma_{p_2}^z
        \sigma_{p_3}^z \sigma_{p_4}^z ,
        \end{split}
    \label{obs_list}
\end{equation}
which measure the nearest neighbor two-spins correlation, nearest neighbor
hopping amplitude, zero-momentum distribution function, and the plaquette
interaction of four spins.  The sum in $O^F$ is over all plaquettes and
$\sigma_{p_i}$~($i=1,2,3,4$) refers to four spins around a plaquette $p$.

\section{Numerical study of eigenstate thermalization
hypothesis}\label{sec:3}

\begin{figure}[t]
    \includegraphics*[width=\columnwidth]{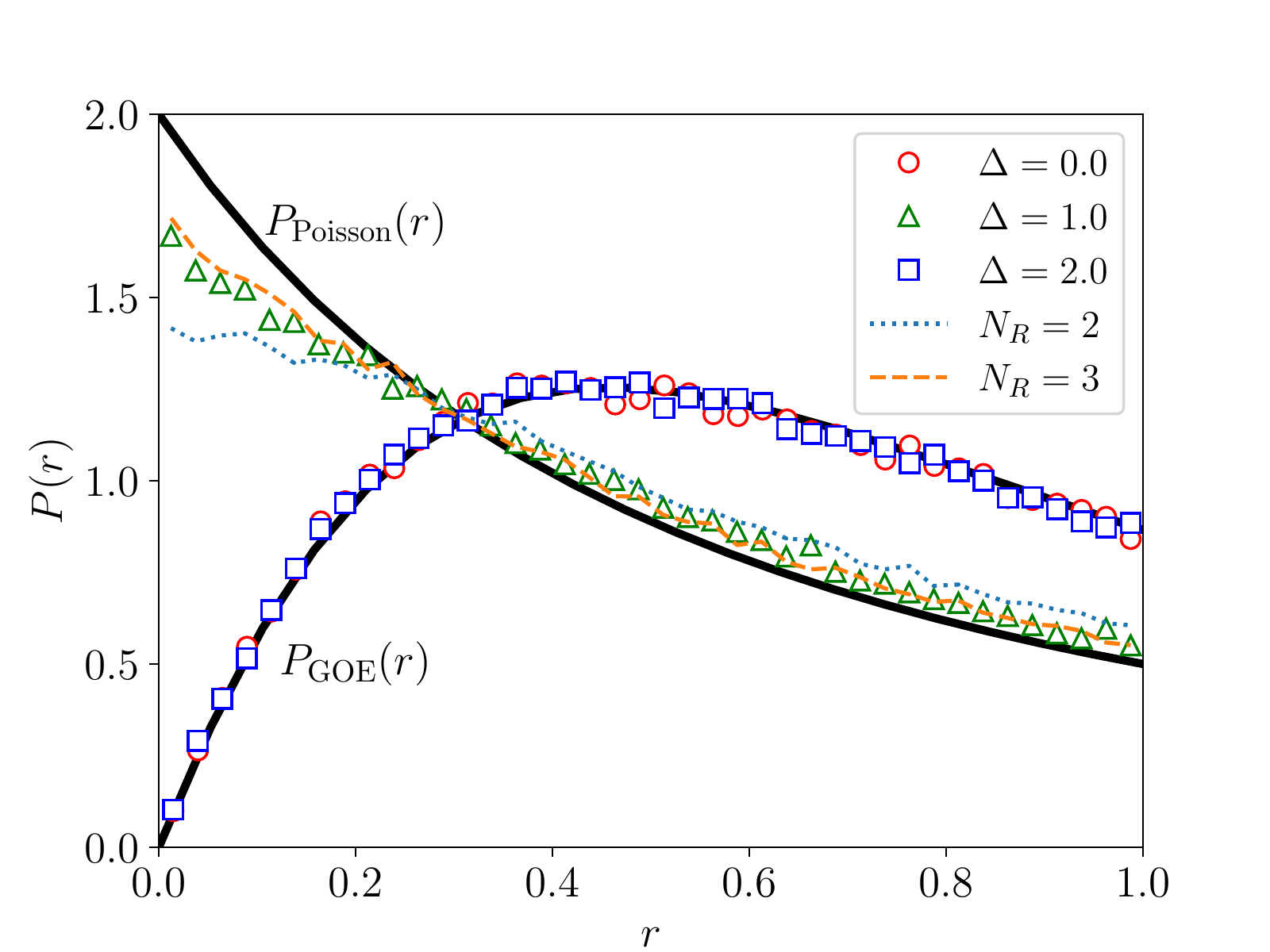}
    \caption{Distributions of the ratio of consecutive energy gaps of the {\em XXZ}
        model with $\lambda=1$ on the rectangular lattice of size $4\times
        7$. These data are obtained from the half of the energy
        eigenvalues in the middle of the entire spectrum.  They are compared
        with the corresponding distribution from the Poisson-distributed
    energy spectrum, $P_{\rm Poisson}(r)$, and the random matrix spectrum in the
    Gaussian orthogonal ensemble, $P_{\rm GOE}(r)$. 
        The peculiar shape of the distribution at $\Delta=1$ 
        is ascribed to the SU(2) symmetry,
        which will be analyzed in detail in Sec.~\ref{sec:su2}.
        The dotted and dashed lines are from a mixture 
        of replicated spectra, which will be also explained in 
        Sec.~\ref{sec:su2}.
    } \label{fig2}
\end{figure}

\subsection{Ratio of consecutive energy gaps}

As a signature for the quantum chaos, we
investigate the statistics of the ratio of consecutive energy
gaps~\cite{Oganesyan.2007,Atas.2013}:
\begin{equation}
    r_n = \min\left[ \frac{E_{n+1}-E_n}{E_n-E_{n-1}} ,
    \frac{E_n-E_{n-1}}{E_{n+1}-E_n} \right] .
    \label{r_def}
\end{equation}
Figure~\ref{fig2} shows the numerical data obtained with the parameters
$\Delta=0, 1$, and $2$ with fixed $\lambda=1$. When $\Delta=0$ and $2$, the
distribution is in good agreement with the distribution function 
\begin{equation}
    P_{\rm GOE}(r) = \frac{27}{4} \frac{(r+r^2)}{(1+r+r^2)^{5/2} },
    \label{Pr_GOE}
\end{equation}
which describes the distribution for random matrices 
in the Gaussian orthogonal ensemble~(GOE)~\cite{Atas.2013}.
The agreement implies that the {\em XXZ} model is quantum chaotic at $\lambda=1$.
We also confirmed the quantum-chaotic behavior at $\lambda=1/2$, which is not shown.

The one-dimensional {\em XXZ} model with $\Delta=0$ and $\lambda=1$ is mapped 
to the free fermion model via the Jordan-Wigner 
transformation~\cite{Lieb.1961}, thus it is integrable. 
The transformation, however, generates nonlocal interaction terms for a 
two-dimensional system. Thus, the two-dimensional {\em XXZ} model is 
nonintegrable even when $\Delta=0$.

At $\Delta=1$, the distribution deviates significantly from 
$P_{\rm GOE}(r)$. It also deviates from 
$P_{\rm Poisson}(r) = 2/(1+r)^2$, which is characteristic
of a nonchaotic system following the Poisson statistics~\cite{Atas.2013}. 
At $\Delta=1$, the system is SU(2) symmetric and the energy eigenvalue 
spectrum in the MSS is a mixture of the spectrum from all the SU(2)
subsectors, which results in a deviation from the GOE 
distribution~\cite{Giraud.2022}.
We will scrutinize the role of the SU(2) symmetry in Sec.~\ref{sec:su2}.

\subsection{Statistics of diagonal elements}\label{sec:diag}
The ETH proposes that matrix elements of an observable $O$, $O_{mn} \equiv
\langle E_m|O|E_n\rangle$, take the form
\begin{equation}
    O_{mn} = g_O(E_{mn}) \delta_{mn} + \frac{e^{-S(E_{mn})/2}}{{N}^{\theta}} 
    f_O(E_{mn},\omega_{mn})
    R_{mn},
    \label{ETH}
\end{equation}
where $E_{mn} = (E_m+E_n)/2$, $\omega_{mn} = (E_m-E_n)$, $S(E)$ is the
thermodynamic entropy~(the Boltzmann constant is
set to unity), $g_O$ and $f_O$ are smooth
functions of their arguments, and $\{R_{mn}\}$ are fluctuating variables
having the statistical properties similar to elements of a random
matrix in the GOE~\cite{Deutsch.1991,Srednicki.1994,D'Alessio.2016}.
The ETH ansatz applies to an operator whose Hilbert-Schmidt norm is
normalized to an $O(1)$ constant~\cite{Schonle.2021}.
The factor ${N}^{-\theta}$ is included in Eq.~\eqref{ETH} as a compensation
for the Hilbert-Schmidt norm of the operators in Eq.~\eqref{obs_list}. 
Specifically, $\theta=1/2$ for $O^{Z,J,F}$ and $\theta=0$ for
$O^P$~\cite{LeBlond.2019,Mierzejewski.2020}.
This ansatz guarantees the quantum thermalization and the
fluctuation-dissipation theorem for isolated quantum 
systems~\cite{D'Alessio.2016,Srednicki.1999,Essler.2012,Nation.2019,Khatami.2013,Noh.2020,Schonle.2021}.
    Note that the quantities $R_{mn}$ follow a Gaussian
    distribution as the random matrix elements in the GOE. 
    We remark, however, that 
    their higher order correlations are not described by the GOE
    random matrix
    theory~\cite{Foini.2019,Brenes.2021,Wang.2022,Dymarsky.2022,Chan.2019,Murthy.2019,Richter.2020}.
    In this work, we focus on the Gaussian nature of the distribution and 
    do not study the higher order correlations.

\begin{figure}[t]
    \includegraphics*[width=\columnwidth]{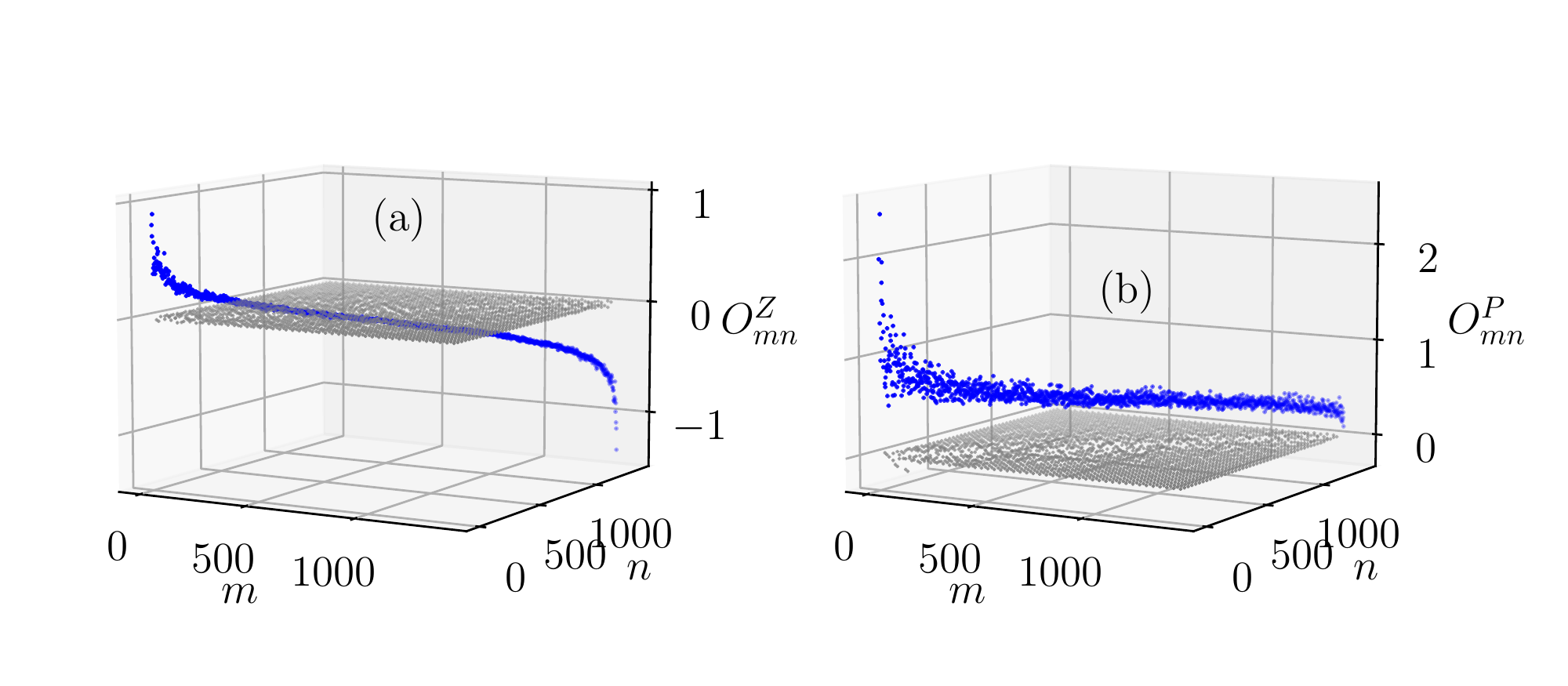}
    \caption{Matrix elements $O^Z_{mn}$ in (a) and $O^P_{mn}$ in (b)
        in the Hamiltonian eigenstate basis with $m,n = 0,\cdots, 1391$. 
        The lattice size is $4\times 5$
        and model parameters are $\Delta = 2$ and $\lambda=1$.
    }\label{fig3}
\end{figure}

Figure~\ref{fig3} presents matrix elements of $O^Z$ and $O^P$.
Diagonal elements, far from the spectrum edges, vary smoothly 
with the energy quantum number.
Offdiagonal elements have a relatively smaller magnitude than  
diagonal elements. 
These overall features are consistent with the ETH ansatz.

\begin{figure}[th]
    \includegraphics*[width=\columnwidth]{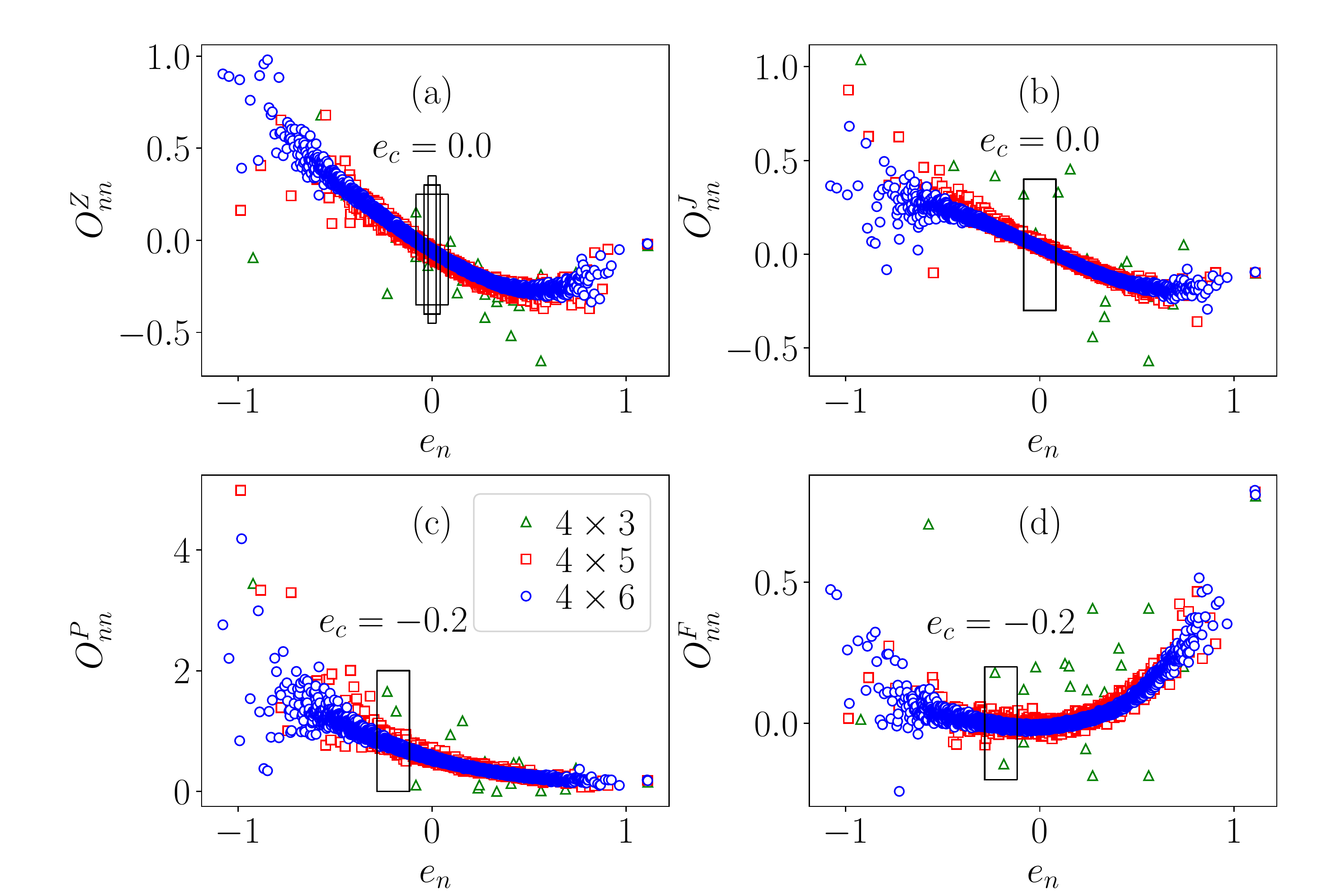}
    \caption{Diagonal matrix elements $O_{nn} = \langle E_n |
        O|E_n\rangle$ versus energy density $e_n = E_n /{N}$ at three
        different lattice sizes with $\Delta=2$ and $\lambda=1/2$. 
        Rectangular boxes represent 
        energy windows $W(e_c N, \delta E)$ of width 
        $\delta E = 0.5, 1.0, 2.0$ for $O^Z$ and $\delta E=2.0$ 
        for the other observables for the system of size $4\times 6$.
    }
    \label{fig4}
\end{figure}

The diagonal elements are plotted in Fig.~\ref{fig4}. 
According to the ETH, diagonal elements $O_{nn}$ 
should follow the Gaussian distribution with mean 
$g_O(E_n)$ and variance $e^{-S(E_n)} |f_O(E_n,0)|^2$. 
This ansatz can be tested with the distribution of the diagonal elements
for energy eigenstates in an energy window $W(E_c, \delta E)$, a set of
energy eigenstate whose energy eigenvalues lie within an interval
$E_c - \delta E \leq E_n \leq E_c + \delta E$. 
Rectangular boxes drawn in Fig.~\ref{fig4}(a) represent the energy
windows of width $\delta E=0.5, 1,$ and $2$ with $e_c = E_c/N = 0.0$.
The distribution of the diagonal elements
within an energy window is influenced by two
factors~\cite{Ikeda.2015lv9,Mierzejewski.2020}:
(i) intrinsic eigenstate-to-eigenstate fluctuations
and (ii) extrinsic fluctuations due to a systematic energy dependence 
of the diagonal elements.  
It is clear that the extrinsic fluctuations become dominant as $\delta E$ 
increases. 

\begin{figure}[th]
    \includegraphics*[width=\columnwidth]{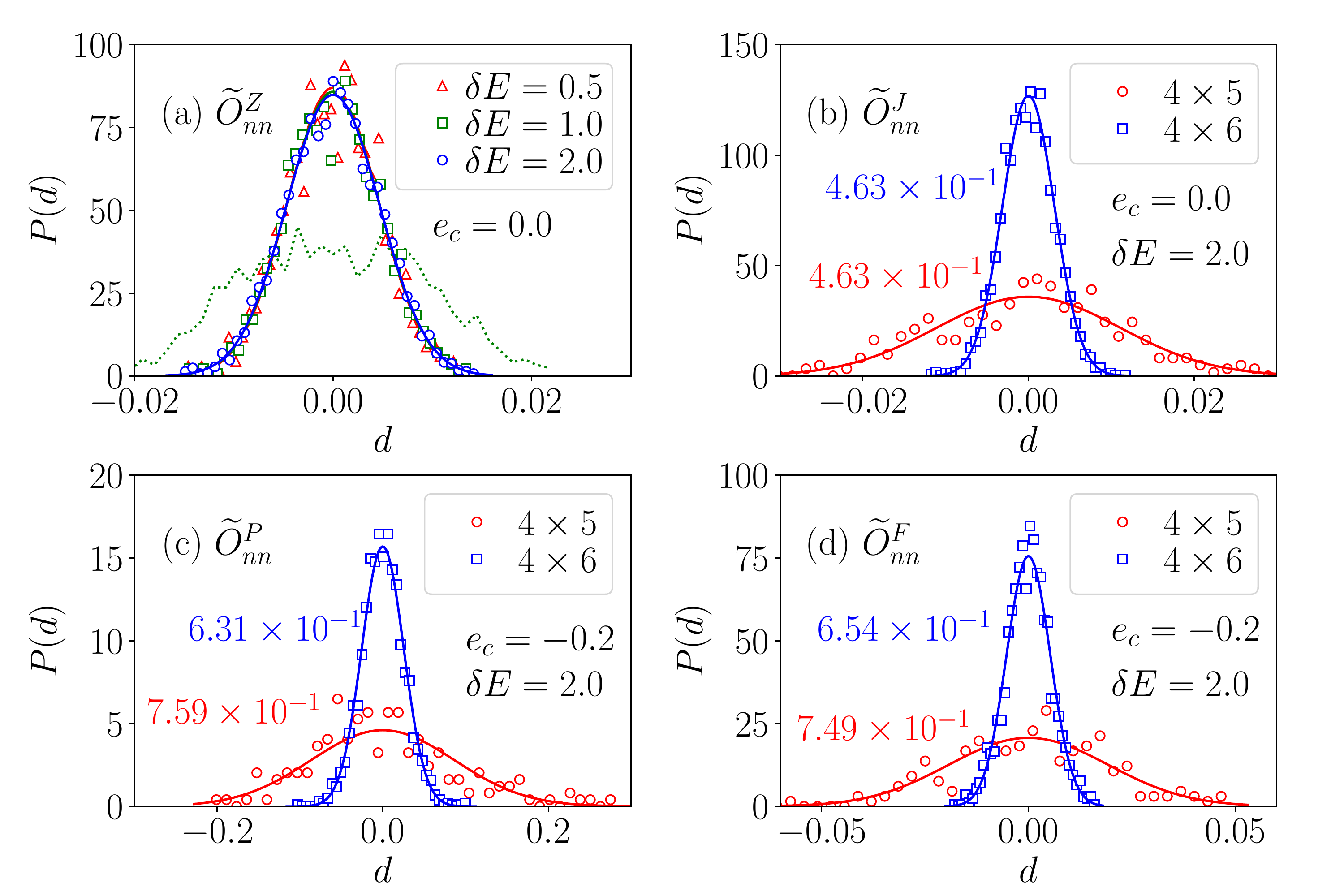}
    \caption{Distribution of detrended diagonal matrix elements
        $d = \widetilde{O}_{nn}$ within the energy window 
        $W(E_c=e_c N, \delta E)$ depicted with the rectangular boxes in
        Fig.~\ref{fig4}.  Model parameters are $\Delta = 2$ and $\lambda=1/2$. 
        (a) We compare the distributions for the operator $O^Z$ 
        with the choice of three different values
        $\delta E$ when the lattice is of size $4\times 6$. 
        The solid curves represent the Gaussian distribution of the same
        mean and variance as the histogram data. The dotted line is the
        probability distribution of the bare diagonal elements, after being
        subtracted by their mean value, with $\delta E=1.0$.
        (b)-(d) We compare the
        distributions obtained from the lattices of size $4\times 5$ and 
        $4\times 6$. All the distributions are consistent with the Gaussian
        distributions (solid curves). 
        Numerical values of $\sigma_d^2 N^{2\theta} D$~(see main text) 
    are annotated in (b)-(d).
    }\label{fig5}
\end{figure}

In order to reduce a finite $\delta E$ effect and 
isolate the intrinsic fluctuations, we
introduce a {\em detrended} diagonal element~\cite{Ikeda.2015lv9}
\begin{equation}
    \widetilde{O}_{nn} = O_{nn} - h_W(E_n) ,
    \label{detrended}
\end{equation}
where $h_W(E)$ is a fitting function to $O_{nn}$ within an energy window
$W(E_c,\delta E)$.
In this work, we choose a linear function for $h_W(E)$.
In Fig.~\ref{fig5}(a), we compare the distributions of the detrended 
diagonal elements of $O^Z$ with three different values of 
$\delta E=0.5, 1.0$, and $2.0$. 
Those distributions are almost identical to each other, 
which implies that the detrending removes 
the extrinsic fluctuations. We also present the distribution 
of the {\em bare} diagonal elements within the energy window of width $\delta
E=1$. They are shifted to have zero mean. The bare distribution is much 
broader than the detrended distribution due to the extrinsic fluctuations. 
This comparison demonstrates that the detrending is useful.
It allows one to take a large value of $\delta E$ for better
statistics without suffering from the finite $\delta E$ effect. 
In Figs.~\ref{fig5}(b)-(d), we present the distributions
of the detrended diagonal elements of the observables $O^J$, $O^P$, and 
$O^F$ within the energy windows shown in Figs.~\ref{fig4}(b)-(d).
The numerical results are in good
agreement with the Gaussian distributions of the same mean and variance,
which supports the ETH. 

According to the ETH in Eq.~\eqref{ETH}, the variance of the diagonal elements 
$\sigma_d^2$ normalized with the system size, $\sigma^2_d N^{2\theta}$, 
should be inversely proportional to the density of states 
$D \simeq |W(e_cN, \delta E)|/\delta E = e^{S(e_c N)}$.  
This scaling law can be checked by using a plot of 
$\sigma^2_d N^{2\theta}$ against $D^{-1}$ for more than three different 
system sizes, as was done in
Ref.~\cite{Noh.2021}. 
In the current work, numerical data are available from only two 
different system sizes $4\times 5$ and $4\times 6$. 
    Due to the limited range of system sizes, we cannot perform such 
    a systematic
    finite size scaling analysis. Alternatively, we only report 
the quantitative values of $\sigma_d^2 N^{2\theta} D$.
The numerical values at two different system sizes, 
shown in Figs.~\ref{fig5}(b)-(d), are close to each other 
up to a relative error of $\lesssim 15\%$, which supports the scaling
behavior $\sigma_d^2 N^{2\theta} \propto 1/D$.

\subsection{Statistics of off diagonal elements}

\begin{figure}[ht]
    \includegraphics*[width=\columnwidth]{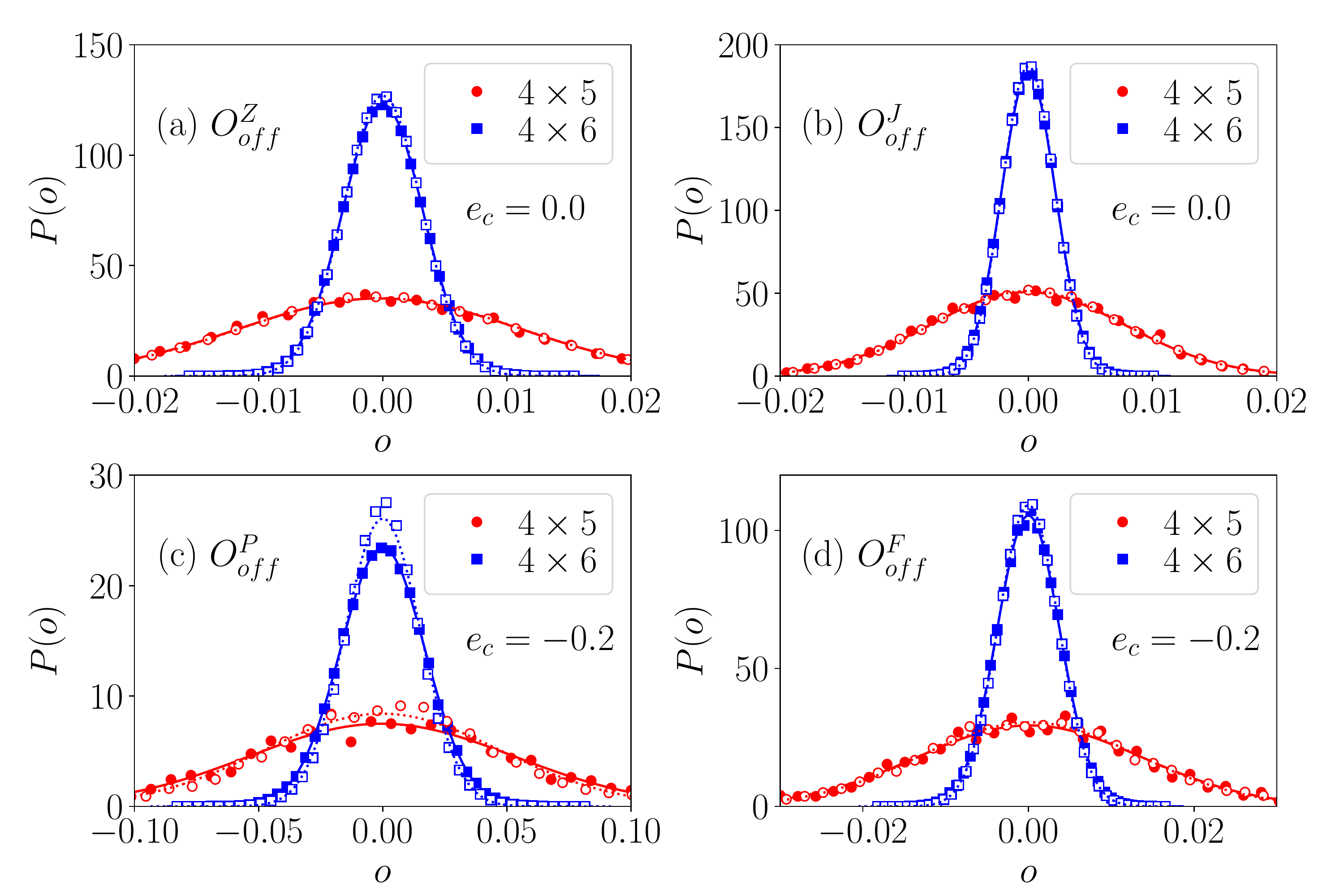}
    \caption{Distribution of offdiagonal matrix elements $O_{mn}$ with
        $m \neq n$ among energy eigenstates within the energy window $W(E_c =
        e_c N, \delta E)$ of width
        $\delta E=0.5$~(filled symbols and solid lines) and $\delta
        E=1.0$~(open symbols and dashed lines) centered at
        the energy density $e_c = 0.0$ or $-0.2$.
        The curves represent the Gaussian distribution with the same mean
        and variance as the histogram data. 
        The model parameters are $\Delta = 2$ and $\lambda=1/2$.
    }\label{fig6}
\end{figure}

We also investigate the statistical property of the offdiagonal elements
$o=O_{mn}$ for $|E_n\rangle$ and $|E_m\rangle \in W(E_c=e_cN, \delta E)$ 
with $n\neq m$. 
    These offdiagonal elements correspond to the term
    $\frac{e^{-S(e_cN)}}{N^\theta} f_O(e_cN, \omega\simeq 0 ) R_{mn}$ 
    with $m\neq n$ in the ETH ansatz of Eq.~\eqref{ETH}.
Figure~\ref{fig6} presents the distributions for the four
observables. 
Each numerical distribution function is in good agreement with the 
Gaussian distribution of the same mean and variance, which is consistent
with the ETH.

\begin{figure}[ht]
    \includegraphics*[width=\columnwidth]{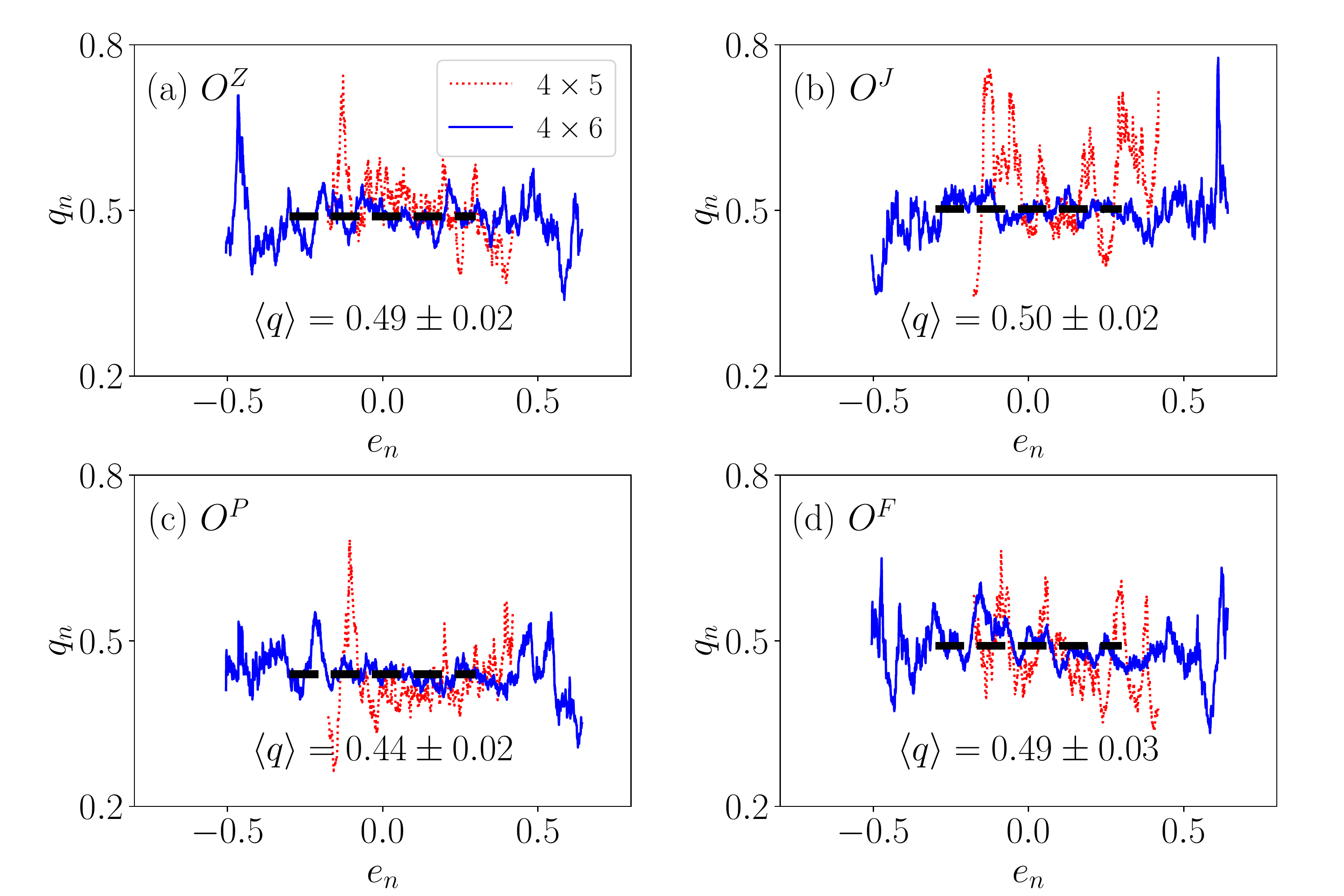}
    \caption{Variance ratio $q_n = \sigma_o^2/\sigma_d^d$ 
        for the model with $\Delta=2$ and $\lambda=0.5$ and $L_1\times L_2=
        4\times 5$~(dotted line) and $4\times 6$~(solid line).
        The mean value and the standard deviation of the ratios $\{q_n\}$
        within the energy interval $-0.3<e_n < 0.3$ are presented 
        in each panel~(broken line) for $L_1\times L_2=4\times 6$.
    }\label{fig7}
\end{figure}

To test the ETH further, we compare the variances $\sigma^2_d$ and
$\sigma^2_o$ of the  diagonal and offdiagonal elements, respectively.
For each energy eigenstate $|E_n\rangle$, we construct 
an energy window $W(E_c=E_n, \delta E)$, calculate the matrix elements,
and evaluate
a variance ratio $q_n = \sigma_o^2 / \sigma_d^2$. The diagonal elements are
detrended as explained in Sec.~\ref{sec:diag}. The ratio $q_n$ obtained with
$\delta E=0.5$ is plotted as a
function of the energy density $e_n = E_n/N$ in Fig.~\ref{fig7}.
The ratio is fluctuating around the mean value, and the amplitude of
fluctuations decreases as the system size 
increases except for the spectrum edges. 
The mean value is close to $1/2$, which is also consistent with the ETH
prediction.

We add a remark on a finite $\delta E$ effect. 
The shape of the distributions shown in Fig.~\ref{fig6} 
varies slightly with $\delta E$.
According to the ETH, an offdiagonal element $O_{mn}$ is a Gaussian 
random variable of variance
$e^{-S(E_{mn})} |f_O(E_{mn},\omega_{mn})|^2/{N}^{2\theta}$. 
Given a finite value
of $\delta E$, the term $e^{-S(E)} |f_O(E,\omega)|^2$ may vary 
around a mean value $e^{-S(E_c)} |f_O(E_c,0)|^2$ up to $O(\delta E)$. 
Unlike the case for diagonal elements, the variation leads to a subleading
contribution to the variance of offdiagonal elements. Thus, a finite-$\delta
E$ effect is weak for the offdiagonal elements. 
The numerical results in Fig.~\ref{fig6} shows that such an effect is
indeed negligible for $O^Z$, $O^J$, $O^F$ with $\delta E=0.5$. 
On the other hand, it is still noticeable for $O^P$ in Fig.~\ref{fig6}(c).
We attribute the result $\langle q\rangle \simeq 0.44$ in Fig.~\ref{fig7}(c)
to a finite-$\delta E$ effect.

\section{SU(2) symmetric {\em XXZ} model with $\Delta=1$}\label{sec:su2}
We have shown that the {\em XXZ} model in the symmetry-resolved MSS obeys the ETH.
When $\Delta =1$, the {\em XXZ} Hamiltonian has an additional symmetry 
under the global spin rotation, SU(2) symmetry. 
Thus, the MSS can be further decomposed into 
the symmetry subsectors, called SU(2) subsectors, each of which is 
characterized with the total spin quantum number 
$s$ as described in Sec.~\ref{sec:2}. 
In this section, we investigate whether the ETH is also valid for 
the SU(2)-symmetric {\em XXZ} model.

In Fig.~\ref{fig2}, we have seen that the distribution $P(r)$ for the
ratio of consecutive energy gaps at $\Delta=1$ deviates from $P_{\rm
GOE}(r)$ and $P_{\rm Poisson}(r)$. The SU(2) symmetry 
is responsible for it. The MSS is the union of the SU(2)-symmetric 
subsectors.  
Recently, it was found that presence of symmetry subsectors modifies
the gap ratio distribution function from the universal 
form~\cite{Giraud.2022}. 
Even if the energy spectrum in each subsector follows the GOE statistics, 
$P(r)$ from the whole spectrum is characterized by a distinct form 
determined by the number of subsectors and their relative
sizes~\cite{Giraud.2022}.

In order to understand the shape of $P(r)$ at $\Delta=1$, 
we construct an artificial set of energy eigenvalues
$\mathcal{E}(N_R)$ as the union of shifted replicas 
of the real energy spectrum $\{E_n\}$ obtained at $\Delta=2$, $\mathcal{E}(N_R) =
\cup_{p=1}^{N_R} \{E_n + (p-1) \Delta E\}$ with $N_R$ the number of replicas.
We took $\Delta E~(=0.1)$ which is much larger than the mean level spacing.
    Figure~\ref{fig2} shows the distribution functions of the superimposed
    spectrum with $N_R=2$~(dotted line) and $N_R=3$~(dashed line).
    These data confirm that the the distribution of the
    superimposed spectrum is different from $P_{\rm Poisson}$ and $P_{\rm
    GOE}$~\cite{Giraud.2022}. We note that the distribution function $P(r)$
    at $\Delta =1$ lies between the distribution functions from 
    the superimposed energy spectrum of $N_R = 2$~(dotted line) or 
    $3$~(dashed line) replicas.
This comparison suggests that the energy spectrum in each SU(2) subsector
obeys the GOE statistics and that a few~$(2\sim 3)$ SU(2) subsectors are
dominant in the MSS.

\begin{figure}[ht]
    \includegraphics*[width=\columnwidth]{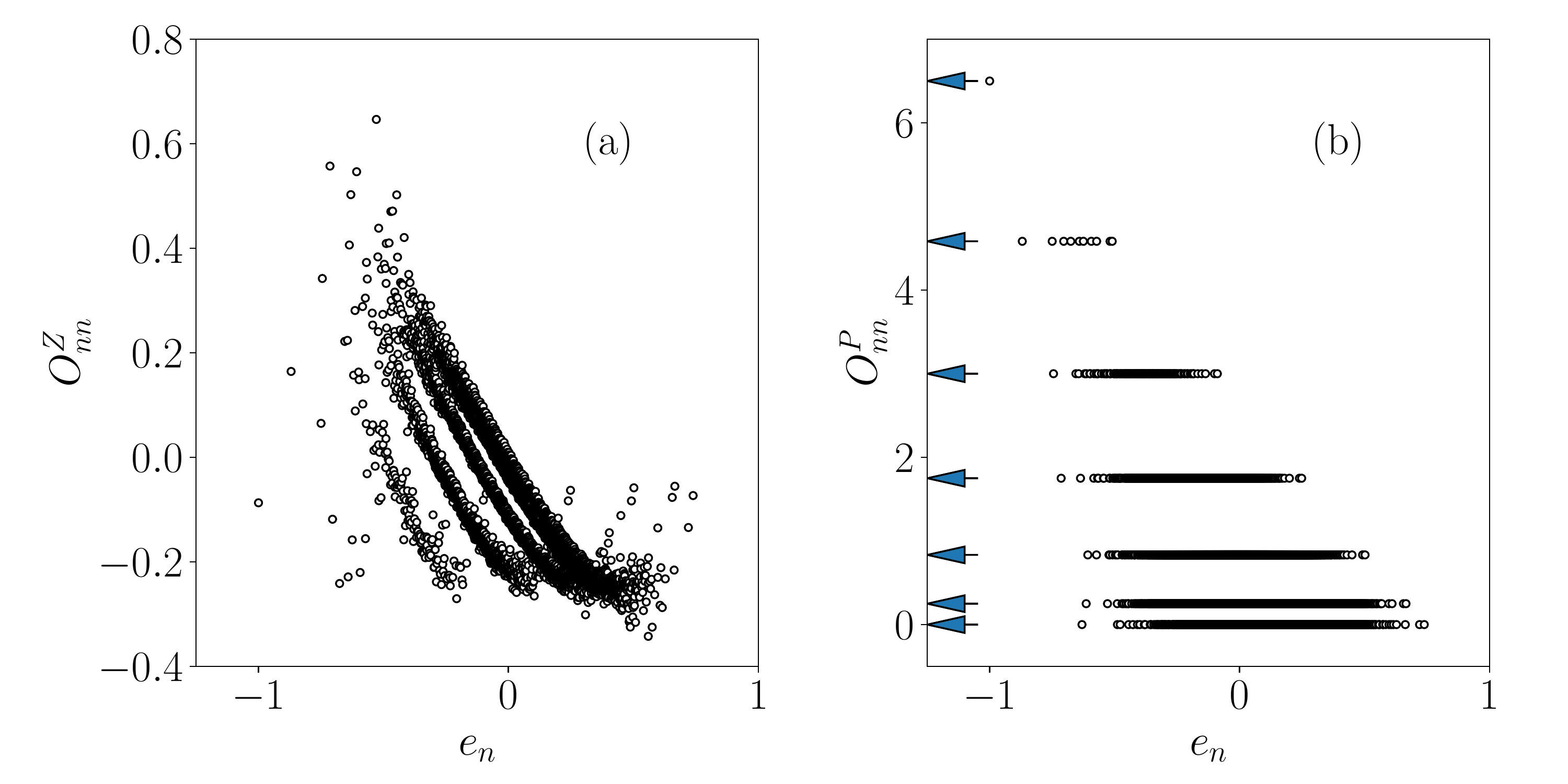}
    \caption{Diagonal elements of the operators $O^Z$ and $O^P$ in the
        energy eigenstate basis when $\Delta = 1$, $\lambda=1/2$, and
        $L_1\times L_2 = 4\times 6$. Note that the diagonal elements of
        $O^P$ are quantized to the values $s(s+1)/N$ with $N=24$ and $s=0,
        2, \cdots, 12$ as indicated by arrows.
    }\label{fig8}
\end{figure}

Figure~\ref{fig8} shows the diagonal elements of the operators $O^Z$ and
$O^P$ in the energy eigenstate basis at the SU(2) symmetric point~($\Delta
=1$ and $\lambda=1/2$). One finds that the diagonal elements are organized
into several branches. Moreover, the diagonal elements of $O^P$ are
quantized.
Note that $O^P$ defined in Eq.~\eqref{obs_list} is rewritten as
\begin{equation}
    O^P = S^+ S^- / N= (\bm{S}^2 - (S^z)^2 + S^z)/N
    \label{P2S}
\end{equation}
in terms of the total spin operator $\bm{S} = \frac{1}{2} 
\sum_{\bm{r}} \bm{\sigma}_{\bm r}$. 
Thus, the diagonal element of $O^P$ in the MSS~[$(S^z)'=0$] takes a
quantized value  
\begin{equation}
    \langle E_n | O^P | E_n \rangle = s_n (s_n + 1) / N
    \label{P2j}
\end{equation}
with a nonnegative integer $s_n$ equal to or less than 
$s_{max} = N/2$.  Since $N=L_1 L_2$ is even in
this work, the total spin quantum number takes an integral value.

\begin{figure}[ht]
    \includegraphics*[width=\columnwidth]{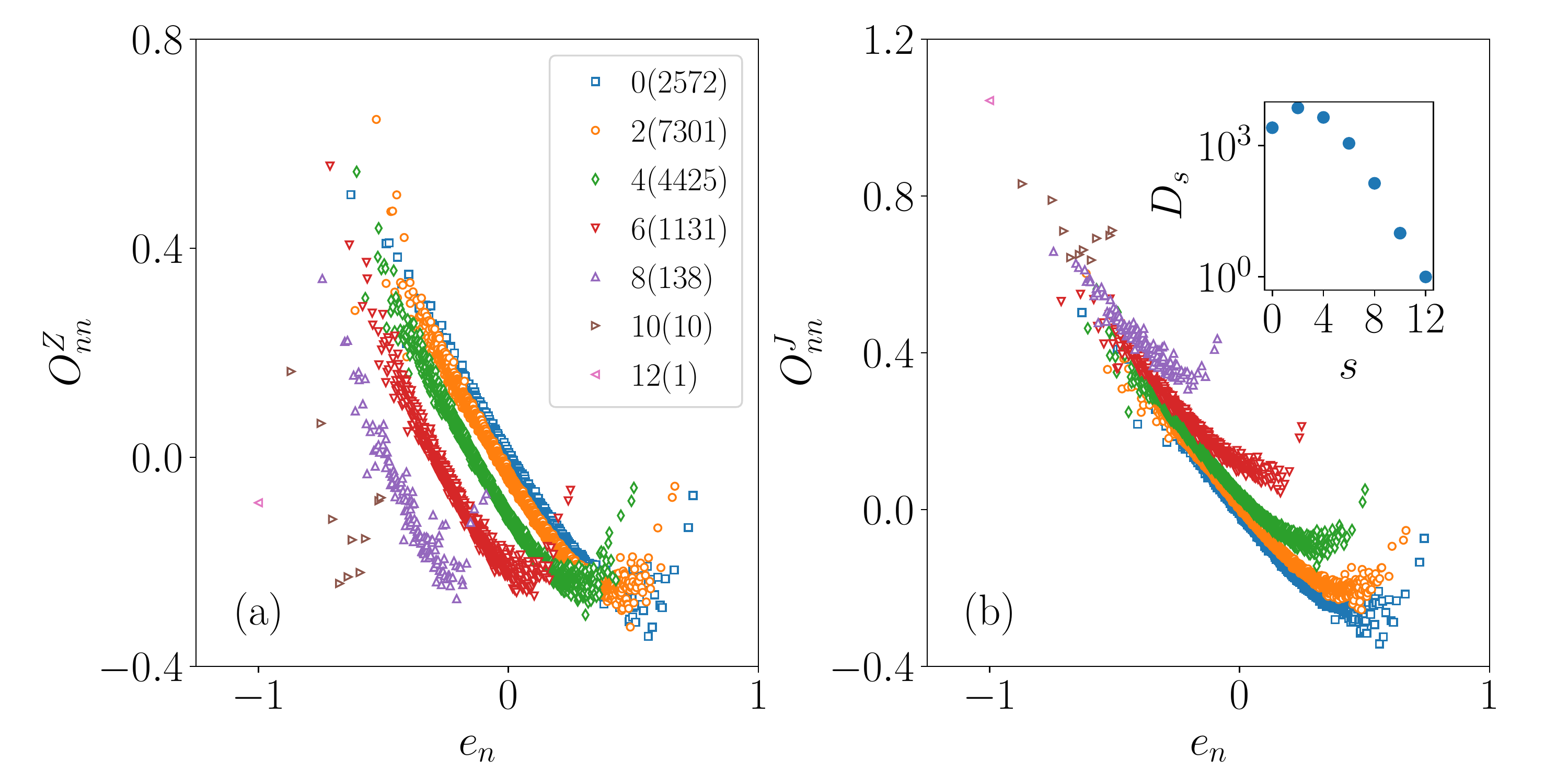}
    \caption{Diagonal elements of the operators (a) $O^Z$ and (b) $O^J$ 
        when $\Delta = 1$, $\lambda=1/2$, and $L_1\times L_2 = 4\times 6$.
        Diagonal elements are plotted with different symbols depending on
        their total spin quantum number $s$. The inset in (b) shows 
        $D_s$, the number of eigenstates in the SU(2) subsector of 
        total spin quantum number $s$.
    }\label{fig9}
\end{figure}

Using the quantization in Eq.~\eqref{P2j}, 
one can identify the total spin quantum number $s$ of an energy eigenstate.
We present the diagonal elements of $O^Z$ and $O^J$
in each SU(2) subsector in Fig.~\ref{fig9}.
It is clear that the branch corresponds to the SU(2) subsector.
Note that the SU(2) subsectors with odd $s$ are missing.
An odd $s$ is not compatible with the other symmetries in the MSS. 

Before proceeding further, we briefly review the theory of spin addition.
Consider two spins $\bm{S}_1$ and $\bm{S}_2$ with
$(\bm{S}_{1,2}^2)' = s_{1,2} (s_{1,2}+1)$. 
The sum of them $\bm{S} = \bm{S}_1 + \bm{S}_2$ has an eigenvalue
$(\bm{S}^2)' = s(s+1)$ where $s = |s_1 - s_2|, |s_1 - s_2|+1, \cdots ,
s_1+s_2$~\cite{Sakurai.2011}. 
Thus, the Hilbert space for the two spins can be represented as a
direct product of the Hilbert space of individual spins or as 
a direct sum of the total spin sectors~\cite{Cirac.1999,Toth.2005,Cohen.2016}:
\begin{equation}
\bm{( 2s_1+1) } \bigotimes \bm{(2 s_2+1)} = \bigoplus_{s=|s_1-s_2|}^{s_1+s_2} 
\bm{(2 s+1)},
    \label{2spinSum}
\end{equation}
where $\bm{(2s+1)}$ stands for a $(2s+1)$-dimensional Hilbert space 
consisting of $(2s+1)$ states characterized by the total spin quantum 
number $s$ and the magnetization quantum number 
$m_z\equiv (S_z)'=-s,-s+1,\cdots,s$.
Applying the addition rule
iteratively, one can find that the Hilbert space for $N$ spin-1/2 particles 
is given by~(assuming that $N$ is even for a notational simplicity) the
Clebsch-Gordan decomposition series
\begin{equation}
    \bm{(2)}^{\bigotimes N}
    = \bigoplus_{s=0}^{N/2} m_{N,s} \bm{(2s+1)},
    \label{CGdecomp}
\end{equation}
where the multiplicity factor $m_{N,s}$ is given by
\begin{equation}
    m_{N,s} = \frac{N! (2s+1)}{\left(\frac{N}{2}-s\right)!
    \left(\frac{N}{2}+s+1\right)!} .
    \label{Catalan}
\end{equation}
%The multiplicity factor $m_{N,s}$ is the number of states characterized by
%the spin quantum number $s$ and the magnetization quantum number $-s\leq
%m_z\leq s$. Using the Stirling formular, we find that 
The multiplicity factor $m_{N,s}$, as a function of $s$, takes a maximum
value at $s = s_{M} \simeq {\sqrt{N}}/{2}$ for large $N$. 

The MSS considered in this work is characterized with $m_z=0$ and the other
symmetry constraints.
Thus, the number of spin-$s$ eigenstates in the MSS, denoted as $D_s$,
is equal to or smaller than $m_{N,s}$. 
It is counted numerically and plotted in Fig.~\ref{fig9}(b). 
It is maximum at $s=2$, which is close to 
the peak position of $m_{N,s}$, $s_M = \sqrt{N}/2\simeq 2.4$ for $N=24$.

\begin{figure}[ht]
\includegraphics*[width=\columnwidth]{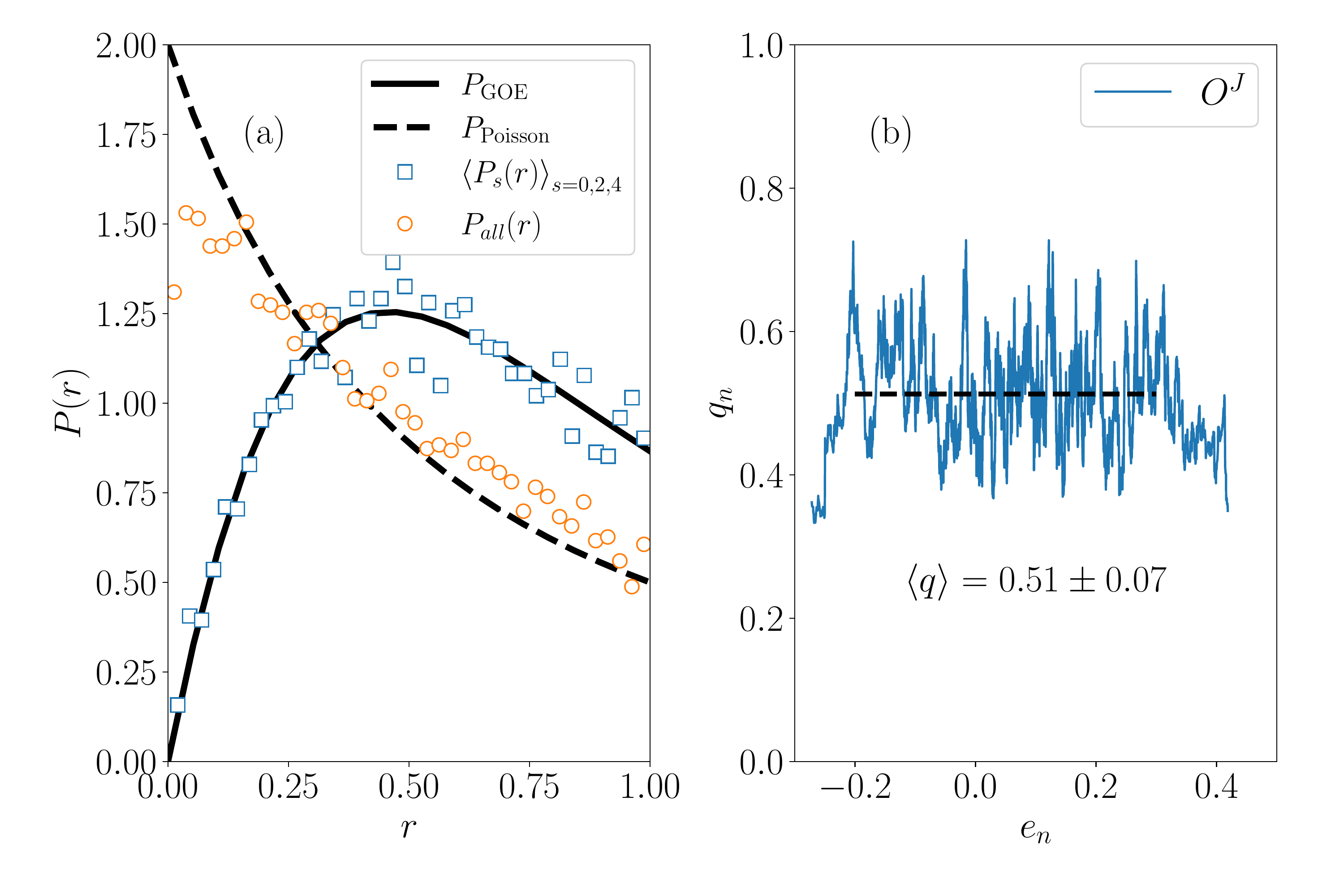}
\caption{(a) Distribution of the ratio of consecutive energy gaps.
    (b) Ratio of the variance of offdiagonal
    elements to the variance of diagonal elements of the operator
    $O^{J}$ in the SU(2) subsector of $s=2$.
    $\langle q\rangle$ denotes the average  of $q_n$ within the energy interval
    $-0.2\le E_n/N \le 0.3$. Both data are obtained from the system with
    $\Delta=1$, $\lambda=1/2$, and $L_1\times L_2 = 4\times 6$.
}\label{fig10}
\end{figure}

It is an intriguing question whether the SU(2) symmetric {\em XXZ} model is still 
quantum chaotic and obeys the ETH.
We focus on the dominant SU(2) subsectors with spin quantum number $s=0, 2, 4$.
We first measure the gap ratio distribution function $P_s(r)$ at each 
SU(2) subsector, and take the average of them 
to evaluate $\langle P_s(r)\rangle_{s=0, 2, 4}$. 
It is in good agreement with $P_{\rm GOE}(r)$,
which indicates that the system is quantum chaotic inside the subsector~[see 
Fig.~\ref{fig10}(a)]. We also measure the distribution function, denoted as
$P_{all}(r)$, using all the energy levels of the subsectors with $s=0, 2, 4$.
It deviates from both $P_{\rm GOE}(r)$ and $P_{\rm Poisson}(r)$ 
as already seen in Fig.~\ref{fig2}.

The ETH ansatz in Eq.~\eqref{ETH} is also tested for the matrix
elements $O_{mn}$ between the energy eigenstates belonging to a single
SU(2) subsector.
We choose the subsector with $s=2$ that contains 
the largest number of eigenstates. 
To a given energy eigenstate $|E_n\rangle$, we construct a similar 
energy window consisting of 101 consecutive energy eigenstates with quantum
numbers from $n-50$ to $n+50$,
calculate matrix elements, and evaluate the variance ratio $q_n
= \sigma_o^2 / {\sigma}_d^2$.
It is plotted in Fig.~\ref{fig10}(b) as
a function of the energy density $e_n = E_n/N$. The ratios far from
the band edges fluctuate around the mean value $\langle q\rangle = 0.51$,
which is close to $1/2$ predicted by the ETH.
The statistical properties of the energy levels and the matrix elements of
observables indicate that the SU(2) symmetric {\em XXZ} model is quantum-chaotic
and obeys the ETH when it is restricted to a total spin-$s$ subsector.

\section{Summary and discussions}\label{sec:summary}
In this paper, we study the statistical properties of the 
energy eigenvalues and eigenvectors of the {\em XXZ} model in 
two-dimensional~(2D) 
rectangular lattices using the numerical exact diagonalization technique.
We showed that the energy eigenvalues spectrum follows the GOE statistics
and that the matrix elements of observables in the energy eigenstate basis
obey the ETH ansatz in the maximum symmetry sector without the SU(2)
symmetry~($\Delta \neq 1$). These results imply that the 2D {\em XXZ} spin
system thermalizes for itself.
The ETH has been tested mostly in one-dimensional systems. There are only a
few works on the transverse-field Ising spin
system in two dimensions~\cite{Fratus.2015,Mondaini.2016,Mondaini.2017}. 
Our work extends the applicability to the 2D {\em XXZ} system which possesses
a larger set of symmetry operators than the Ising system.

When the spin-spin interaction is isotropic~($\Delta = 1$), the {\em XXZ}
Hamiltonian is SU(2) symmetric and the total spin $s$ is a good 
quantum number. The MSS is further decomposed as the direct sum of
SU(2) subsectors.
The SU(2) symmetry modifies statistical
properties of the Hamiltonian eigenspectrum: (i) The energy gap ratio
distribution $P(r)$ deviates from the GOE distribution~(see
Fig.~\ref{fig2}).  (ii) The matrix elements of observables in the energy
eigenstate basis are organized into distinct branches~(see Fig.~\ref{fig8}). 
We showed that these features originate from the emergence of the
subsectors. $P(r)$ deviates from the GOE distribution because the energy
spectrum is a mixture of energy eigenvalues from the subsectors.
Each branch in Fig.~\ref{fig8} corresponds to a total spin 
subsector~(see Fig.~\ref{fig9}). We also showed that the SU(2) symmetric {\em XXZ}
model is still quantum chaotic and satisfies the ETH when it is restricted
to a subsector with a definite spin quantum number.

The SU(2) symmetry raises an intriguing question about the 
thermal equilibrium state. 
The ETH guarantees that an isolated quantum system in an initial state
$|\Psi(0)\rangle$ with an energy expectation value $E$ thermalizes 
in the sense that $\lim_{t\to\infty} \langle\Psi(t)|O|\Psi(t)\rangle = {\rm
Tr} \rho_{eq} O$ for a local observable $O$ with the thermal equilibrium 
density operator $\rho_{eq}$. It can be the microcanonical ensemble state
$\rho_{mc}(E) = \frac{1}{\Omega(E)} \sum_{|E_n-E|<\Delta E} 
|E_n\rangle\langle E_n|$ or the canonical ensemble state $\rho_{c}(\beta) =
\frac{1}{Z(\beta)} e^{-\beta H}$ with the inverse temperature $\beta$
determined by the condition $E = {\rm Tr} \rho_c(\beta) H$. 
Thus, when the initial
state falls in a SU(2) sector with a definite quantum number $(s, m_z)$, the
equilibrium state will be described by the microcanonical or canonical 
ensemble state projected to the SU(2) sector of $(s,m_z)$, denoted as
$\rho_{mc}(E;s,m_z)$ or $\rho_c(\beta; s, m_z)$, respectively.

We can infer the thermal equilibrium state for a state whose total spin 
is distributed around a mean value $\langle \bm{S}^2\rangle$ while  
the magnetization $m_z$ is a good quantum number.
The logarithm of the multiplicity factor $m_{N,s}$ in Eq.~\eqref{Catalan} 
is a concave function of $s$, i.e., $m_{N,s} \ge \sqrt{m_{N,s-1}
m_{N,s+1}}$. Thus, one can generalize the canonical ensemble state to the
grand canonical ensemble-type state
\begin{equation}
    \rho_{g}(\beta,\mu_s; m_z) = 
    \frac{1}{Z(\beta,\mu_s)}e^{-\beta H - \mu_s \bm{S}^2}
    \label{rho_grand}
\end{equation}
projected to the
magnetization $m_z$ sector. The chemical potential $\mu_s$ is determined by
the condition $\langle \bm{S}^2\rangle = {\rm Tr} \rho(\beta,\mu_s) \bm{S}^2$.

It is a challenging question of whether a SU(2) symmetric system, 
which is prepared in a state which is not an eigenstate of 
${\bm S}^2$ and $S^z$, thermalizes.
The SU(2) symmetry results in a degenerate Hamiltonian eigenstate spectrum.
If $|n;m_z\rangle$ is a simultaneous eigenstate of the Hamiltonian and
$S^z$, so is $S^{\pm}|n;m_z\rangle$ with the same energy eigenvalue.
The Wigner-Eckart theorem~\cite{Sakurai.2011} imposes a definite relation
among matrix elements of an observable. These features are not common in the
systems obeying the ETH. In addition, the 
magnetization operators $S^x$, $S^y$, and $S^z$ are the conserved
quantities, but they are not commuting mutually. The non-Abelian nature
prohibits a microcanonical ensemble in which the three magnetizations are
specified simultaneously. These features make it hard to predict the proper
thermal equilibrium state and call for a theory generalizing the ETH.
Recently, the non-Abelian thermal state and the non-Abelian eigenstate
thermalization hypothesis have been proposed as a 
remedy for statistical mechanics for the systems with non-Abelian symmetry, 
such as SU(2)~\cite{Halpern.2016,Halpern.2020,Murthy.2022}.
It will be interesting to simulate the time evolution of the SU(2) 
symmetric {\em XXZ} system, prepared in a general state, and investigate the statistical
ensemble, if any, describing the equilibrium state.
We will leave it for a future work.

\begin{acknowledgments}
  This work is supported by a National Research Foundation of Korea~(KRF)
  grant funded by the Korea government~(MSIP)~(Grant No. 2019R1A2C1009628).
\end{acknowledgments}

\bibliography{paper}

\end{document}